\documentclass[%twocolumn,
  preprintnumbers,amsmath,amssymb]{revtex4}
  \usepackage{graphicx}
\begin{document}
\title{Intrinsic ratchets: A Hamiltonian approach}
\author{A.V. Plyukhin}
\email{aplyukhin@anselm.edu}
 \affiliation{ Department of Mathematics,
Saint Anselm College, Manchester, New Hampshire 03102, USA 
}

\date{\today}% It is always \today, today,

\begin{abstract}
An asymmetric Brownian particle subjected
to an external time-dependent force may acquire a net drift velocity,
and thus operate as a motor or ratchet, even if
the external force is represented by an unbiased time-periodic 
%(e.g., harmonically oscillating)
function or by a zero-centered noise. For an  adequate description of such
ratchets,  a conventional Langevin equation linear in the particle's velocity
is insufficient, and one needs to take into account the first nonlinear
correction to the dissipation force which emerges beyond the weak coupling
limit. We derived microscopically the relevant nonlinear Langevin equation
by extending the standard  projection operation technique beyond the weak
coupling limit.
The particle is modeled as a rigid cluster of atoms
and its asymmetry may be geometrical, compositional (when a cluster is
composed of atoms of different types), or due to a combination of both factors. 
The drift velocity is quadratic in the 
external force's amplitude and  increases with decreasing the force's
frequency (for a periodic force) and inverse correlation time
(for a fluctuating force).
The  maximum value of the drift velocity is independent on the particle's
mass and  achieved  in the  adiabatic limit, i.e. for an  infinitely
slow change of the external field. 
\end{abstract}

%\pacs{05.40.-a, 66.10.C-, 82.70.Dd}

\maketitle

\section{Introduction} 
Intrinsic ratchets, first studied and christened in Ref.~\cite{Broeck},
are Brownian motors based on 
an asymmetric Brownian particle subjected to
a time-dependent unbiased external force $F_{ex}(t)$. 
The latter may be either deterministic  (e.g.,  sinusoidally-varying), or noisy with zero mean.
An orientation of the particle with respect to
the direction of $F_{ex}$ is assumed to be fixed, 
which implies that the particle 
moves in one dimension, perhaps along a track or channel.
The latter serves merely as
a geometric constraint and, unlike models of flashing and rocking
ratchets~\cite{review1,review2}, 
does not impose a tilted periodic potential
in order to break the spatial symmetry of the system. 
For intrinsic ratchets the spatial symmetry is broken  by the (intrinsic)
asymmetry of the particle itself. On the other hand, 
the external time-dependent force $F_{ex}(t)$  prevents the particle
from reaching thermal equilibrium with the surrounding thermal bath and
thus breaks the microscopic dynamical symmetry of detailed balance. 
For granular ratchets (not considered here) which interact with molecules
via  dissipative collisions, and therefore by construction are in a
nonequilibrium state, the external force is not
required~\cite{granular1,granular2}.

Considering minimalistic requirements (a single isothermal bath, 
no need for an external spatially periodic and/or biased potential) and 
the omnipresence of fluctuating electric forces (particularly thermal electric noise
in living cells~\cite{Fay,Liu}) which may serve as an external drive $F_{ex}(t)$,
it is tempting to think of intrinsic
ratchets as one of the closest approximations
to  a perpetual motion machine of the second kind that Nature allows.

% SHOW that for standing wave (equilibrium radiation) there is no drift

%If collisions of an asymmetric Brownian object with bath molecules are
%inelastic (we do not consider this case here), 

Besides satisfying conditions of non-equilibrium and spatial asymmetry,
it is generally believed that any  machine  rectifying  thermal fluctuations must  operate
in a nonlinear regime~\cite{review1,review2}. Again, in contrast to many other models,  
the nonlinearity of intrinsic ratchets originates not from 
an external potential but rather from a nonlinear correction to a dissipation force exerted
by the thermal bath on a Brownian particle. In this sense 
intrinsic ratchets are ``intrinsic'' 
not only because of geometric asymmetry of the particle but also because of the inherent
dissipative nonlinearity of Brownian motion.
Clearly,  a conventional linear Langevin equation for the particle momentum $P$
\begin{eqnarray}
  \frac{d}{dt} P(t)=-\gamma_0\, P(t)+F_{ex}(t)+\xi(t)
  \label{LE_lin}
\end{eqnarray}
cannot produce a long-lived average drift 
$\langle P(t)\rangle$ 
if the external force $F_{ex}(t)$ is unbiased 
and the thermal noise force $\xi(t)$ exerted by the bath is zero-centered. 
On the other hand, the microscopic theory of 
Brownian motion teaches us that the linear Langevin equation (\ref{LE_lin})
is actually an approximation 
which may be insufficient in certain cases. 
Namely, equation (\ref{LE_lin})
can be derived, under certain assumptions,  
from the underlying microscopic dynamics
in the  weak coupling limit, i.e.
in leading order in a small mass ratio parameter 
$\lambda=\sqrt{m/M}\ll 1$,
where $m$ is the mass of a molecule of the bath  and $M$ is that of the
particle, see e.g.~\cite{KO}.   
To higher orders in $\lambda$, additional forces nonlinear
in the particle momentum emerge in the Langevin equation~\cite{KO,Mori,VKO,PS2,PF}.
Though typically small, these nonlinear dissipation forces
may lead to new physical effects which do not show up in the weak coupling limit~\cite{Broeck,PS2,PF,Gelin}.
Such effects, the drift of intrinsic ratchets being one
of them, originate technically from 
the coupling of the first moment $\langle P(t)\rangle$ of the particle's momentum
to the second $\langle P^2(t)\rangle$ and/or higher moments.
On the other hand, the linear Langevin equation (\ref{LE_lin}) implies a closed equation for $\langle P(t)\rangle$ and is clearly insufficient.

The coupling of the first and  higher moments of a targeted variable is the most distinctive feature  of stochastic dynamics beyond the weak  coupling limit.  
It can be addressed with an approach based on
either the Langevin~\cite{PS2,PF,Plyukhin_therm} or 
master equation~\cite{Meurs,Broeck2},  the latter is closely related to the $1/\Omega$ expansion method 
of van Kampen~\cite{Kampen_book,Kampen,Plyukhin_physa}.
While being approximate, the description of classical Langevin dynamics beyond the weak coupling limit has proved in many studies to be accurate  and consistent~\cite{Broeck,VKO,PS2,PF,Gelin,Meurs,Kawai,Kawai2,Kawai3,Broeck2}. In particular, it enjoys  thermalization
toward a correct equilibrium distribution at any order of perturbation theory~\cite{Kampen,Plyukhin_physa,Plyukhin_therm}. %The corresponding quantum theory is  more subtle and is not %considered here.
An alternative approach to the same class of problems is developed
in~\cite{Sekimoto,Sasa}.

We shall show that 
for an asymmetric Brownian particle in a homogeneous
thermal bath, by going one order of $\lambda$  
higher than the weak coupling limit, one obtains, instead of (\ref{LE_lin}),
a  nonlinear Langevin equation 
\begin{eqnarray}
  \frac{d}{dt}P(t)=-\gamma_0\, P(t)-\gamma_1\big(P^2(t)-\langle P^2\rangle_e\big)
  +F_{ex}(t)+\xi(t), \label{LE}
\end{eqnarray}
where 
$\langle P^2\rangle_e$ is the equilibrium second moment 
of the particle momentum, and $\xi(t)$ is a zero-centered noise.
The corresponding equation for the first moment reads 
\begin{eqnarray}
  \frac{d}{dt}\langle P(t)\rangle =-\gamma_0\, \langle P(t)\rangle-\gamma_1\big[\langle P^2(t)\rangle-\langle P^2\rangle_e\big]
  +F_{ex}(t). \label{moments1}
\end{eqnarray}
The second moment appears here 
multiplied by the coefficient $\gamma_1$ 
which is of higher order in $\lambda$. 
It is therefore sufficient to complement Eq.  (\ref{moments1}) with the equation for the second moment in leading order in $\lambda$
\begin{eqnarray}
  \frac{d}{dt}\langle P^2(t)\rangle =-2\gamma_0\,\big[\langle P^2(t)\rangle-\langle P^2\rangle_e\big]
  +2\,F_{ex}(t)\,\langle P(t)\rangle, \label{moments2}
\end{eqnarray}
which can be derived from the linear Langevin equation (\ref{LE_lin}), see section IV for details.
It was shown in~\cite{Broeck} that the system of coupled 
equations (\ref{moments1}) and (\ref{moments2})
predicts a systematic
drift of the particle 
even if the external force $F_{ex}(t)$ is time-periodic and unbiased.
For example, for a harmonic drive
$F_{ex}(t)=F_0\sin(\omega t)$ a calculation based on 
Eqs. (\ref{moments1}) and (\ref{moments2}) gives
the average momentum 
$\langle P(t)\rangle$ which oscillates  
with the frequency $\omega$ in such a manner that its positive and
negative semi-periods 
do not completely compensate each other, see the inset in Fig. 3 below. 
As a result,  the net time-averaged particle's momentum and velocity 
does not vanish. A systematic drift 
also may take place if the external 
force $F_{ex}(t)$ is not a regular time-periodic function 
but a  zero-centered nonequilibrium noise.
A key parameter which determines a direction and
magnitude of the drift is the nonlinear dissipation coefficient $\gamma_1$
in (\ref{LE}).
For a symmetric Brownian particle in a uniform thermal bath 
$\gamma_1$ vanishes and so does the drift.

In Ref.~\cite{Broeck},  
the nonlinear term 
$-\gamma_1\big[
\langle P^2(t)\rangle -\langle P^2\rangle_e\big]$
was introduced  in Eq. (\ref{moments1}) phenomenologically
based on an intuitively appealing
requirement that at 
low perturbation order 
the  coupling of the first two moments $\langle P(t)\rangle$ and 
$\langle P^2(t)\rangle $  must be linear and vanish 
in equilibrium. 
In Refs.~\cite{Kawai,Kawai2,Broeck2}, equations (\ref{moments1})  and (\ref{moments2}) 
were derived  kinetically 
within a mesoscopic model where the ratchet is modeled as 
a structureless asymmetric Brownian object interacting with
molecules of the bath  via elastic  collisions.  Explicit expressions for
the coefficient $\gamma_1$ were derived for
two~\cite{Kawai,Kawai2} and three-dimensional~\cite{Broeck2} convex-shaped
Brownian objects. While providing an important insight, these model
calculations are based on specific assumptions that the thermal bath is
an ideal gas of non-interacting molecules, and that 
collisions of molecules with a ratchet are instantaneous, binary, and
uncorrelated.

The main goal of the present paper is to
derive the nonlinear Langevin equation (\ref{LE}) for an asymmetric
Brownian particle
microscopically from the underlying Hamiltonian dynamics. We shall do this 
with a standard projection operator technique extended in two ways. First,
we model a Brownian particle  not as a point-like object (which of course
cannot be asymmetric), but rather as a cluster
of atoms connected by rigid bonds and interacting with molecules of the
surrounding bath
via spherically symmetric and short-ranged potentials.
%The cluster's asymmetry may be  either
%compositional 
%(a cluster composed of different atoms), or 
%geometrical (a cluster of asymmetric shape).
Second, we shall
go one perturbation order
higher than the standard weak coupling 
approximation, which is 
of order $\lambda^2$, retaining terms up to order $\lambda^3$.
The outcome of such derivation 
will be the Langevin equation (\ref{LE}) with the dissipation coefficients
$\gamma_0$ and $\gamma_1$
expressed in terms of microscopic correlation functions. The expression 
for the nonlinear dissipation coefficient $\gamma_1$
qualifies as a new  
fluctuation-dissipation relation, 
additional to the conventional one
for the linear dissipation coefficient $\gamma_0$.

One advantage of such a microscopic approach
is that it offers a more natural framework 
to describe ratchets based on 
nanoscale molecular systems (proteins, nucleic acids, lipids, molecular
assemblies, etc.) whose asymmetry is often 
not merely geometric but due to  the
inherent structural inhomogeneity.
Although in this paper we use the approximation of rigid bonds and do not
consider the internal dynamics of ratchets, it appears that the theory can
be readily extended in that 
direction too.
Another benefit of a microscopic approach is that it shows that the nonlinear
Langevin equation (\ref{LE}) 
is quite generic and does not imply specific assumptions 
and restrictions (e.g., of a thermal bath made of an ideal gas)
typically imposed in model calculations.
Expressing the nonlinear dissipation 
coefficient $\gamma_1$ in terms of a microscopic  correlation function
(rather than specific parameters of a particular system) allows one to
analyze the problem within a more general and unifying framework 
based on fluctuation-dissipation relations.
The price for this generality
is that relevant correlation functions 
are hard to evaluate analytically and to be determined from experiment or
simulation.

Besides a microscopic derivation of nonlinear Langevin equation (\ref{LE})
addressed  in sections II and III, we shall  also exploit this equation to
evaluate the net drift of intrinsic ratchets activated by harmonic and
fluctuating external forces in section IV. 
Results of a few illustrative molecular dynamics simulations of the ratchets
based on simple atomic clusters (a dimer and trimer) are  presented in
section V. Summarizing remarks
are collected in section VI.

\section{Model}

We model an asymmetric  Brownian particle
as a cluster 
of rigidly connected material points, referred below as ``atoms",
enumerated by index $\nu$, and having masses $M_\nu$.
We shall use the terms "particle" and "cluster" interchangeably.
The total mass of the cluster $M=\sum_\nu M_\nu$ is much larger
than the mass $m$ of a molecule of the surrounding thermal bath.
%The the total mass of the particle $M=\sum_\nu M_\nu$ is much larger
%than the mass $m$ of a molecule of the surrounding bath.
%Each atom of the particle interacts  with molecules of 
%the bath through a spherically symmetrical potential. 
The particle's asymmetry may be  geometrical 
(related to the cluster's shape), 
structural (when the cluster is composed of different atoms), see Fig. 1,
or due to a combination of both factors. While each atom is assumed to
interact with bath molecules through  a spherically symmetric potential, 
equipotential surfaces of the
total potential created by all atoms of a cluster may, of course,
lack any symmetry.

The position vectors of particle's 
atoms ${\bf R}_\nu(t)$ are convenient to express as 
\begin{eqnarray}
{\bf R}_\nu(t)={\bf R}(t)+{\bf a}_\nu,
\label{atom_positions}
\end{eqnarray} 
where ${\bf R}(t)=\sum_\nu M_\nu{\bf R}_\nu(t)/M$  is the  position vector 
of the particle's center of mass, and 
${\bf a}_\nu$ are position vectors 
of the atoms 
in the center-of-mass reference frame. 
The particle is constrained to move  
along the $x$-axis keeping fixed its shape and  orientation, the former due
to the rigidity of bonds, the latter due to being attached to an ideal  track,
or several parallel tracks, inducing no friction. 
Accordingly,  the vectors
 $\{{\bf a}_\nu\}$, as well as components $Y$ and $Z$ 
of the center-of-mass position vector ${\bf R}=(X,Y,Z)$ are fixed and do not
change with  time.  Then  the particle's motion is characterized by 
 a single conjugate
 coordinate-momentum pair, namely 
 the $x$-component of  the center-of-mass position 
vector ${\bf R}(t)$   
and the total momentum of the cluster,
\begin{eqnarray}
X=\sum_\nu \frac{M_\nu\,X_\nu}{M}, \qquad P=\sum_\nu M_\nu\,\dot X_\nu. 
\end{eqnarray}
The problem is therefore formally equivalent
to that of a point-like Brownian particle moving in one dimension and
interacting with the bath via 
an asymmetric effective potential.

\begin{figure*}[t]
\includegraphics[height=3cm]{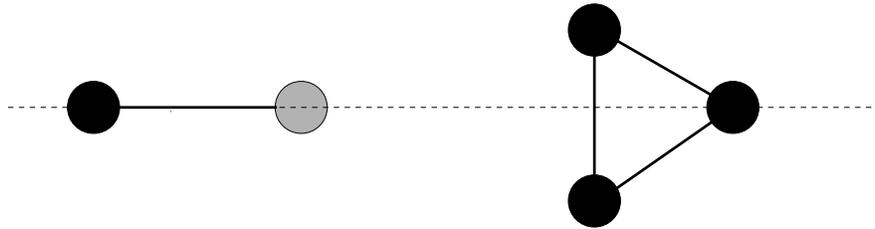}
\caption{Simplest atomic clusters with different types of asymmetry.
  Left: a dimer made of two different atoms (structural asymmetry). 
Right: a trimer made of three identical atoms (geometrical asymmetry). 
%The clusters are constrained to move horizontally
%and without  change of orientation.
}
\end{figure*}

%In what follows we shall refer
%to $P$ as the momentum of the particle and seek for it the
%$Langevin-like equation.

The overall Hamiltonian
of the particle and the thermal bath  has the form
\begin{eqnarray}
H=\frac{P^2}{2M}+H_0+U_{ex}(t).
\label{H1}
\end{eqnarray}
The external potential $U_{ex}(t)$ is assumed to act on each 
atom of the particle, but not on molecules of the bath,
 \begin{eqnarray}
U_{ex}(t)=\sum_\nu u_{ex}({\bf R}_\nu,t)=
\sum_\nu u_{ex}({\bf R}+{\bf a}_\nu,t).
\label{H2}
\end{eqnarray}
The term $H_0=H_0(X)$ is the Hamiltonian of bath molecules in the 
the potential of the Brownian particle when 
the center of mass of the former
held fixed at the position with  $x$-coordinate equal to $X$, 
\begin{eqnarray}
  H_0({X})&=&\sum_i\frac{{\bf p}_i^2}{2m}+V
  +\Phi(X),
\label{H3}
\end{eqnarray}
where $V=\sum_{i>i'} v({\bf r}_i - {\bf r}_{i'})$
is the 
potential for interaction of bath molecules, and
\begin{eqnarray}
  \Phi(X)=\sum_{\nu,i} \phi_\nu({\bf R}_\nu- {\bf r}_{i})=\sum_{\nu, i}
  \phi_\nu({\bf R}+{\bf a}_\nu- {\bf r}_i)
  \label{Phi}
\end{eqnarray}
is the 
potential for interaction of bath molecules and  atoms 
of the particle when the center-of-mass of the latter has position
${\bf R}=(X,Y,Z)$.
In the above  expressions, $\{{\bf r}_i, {\bf p}_i\}$ denote position
vectors and momenta of 
bath molecules,  the subscript $i$ refers
to bath molecules and $\nu$ to atoms of the particle.
The notation $\phi_\nu$ implies that different atoms of the cluster
may interact with the bath via different potentials.
The potentials $v$ and $\phi_\nu$ are assumed to be spherically
symmetric and short-ranged.

%The condition of geometric symmetry of the particle is the  equality 
%$\sum_\nu {\bf a}_\nu=0$, but what is relevant for our model is only the %mirror symmetry along the direction of motion. The corresponding condition %involves only
% $x$-components of vectors 
% ${\bf a}_\nu=(a_{x\nu}, a_{y\nu}, a_{z\nu})$,
%\begin{eqnarray}
%  \sum_\nu a_{x\nu}=0.
%\label{sym_geom}
%\end{eqnarray}
%The condition of structural symmetry is that
%potentials $\phi_\nu$ in (\ref{Phi}) for the interaction with the bath and atomic masses
%are the same for all atoms of the particle, 
%\begin{eqnarray}
%  \phi_\nu=\phi_{\nu'}, \quad M_\nu=M_{\nu'},
%  \qquad \forall \, \nu,\nu'.
%\label{sym_comp}
%\end{eqnarray}
%In what follows, we consider an asymmetric 
%particle for which at least one of the above symmetry conditions does not hold.
%(More generally, the particle is also asymmetric when  its atoms interact %with the external field with different strengths (have different charges), but since in our model
%the atoms are connected by rigid bonds, and
%the external force is applied to the center of mass, 
%this form of asymmetry is irrelevant). 

The Liouville operator $L=\{\cdots, \, H\}$ of the closed system "the particle plus bath"
is a Poisson bracket with the overall Hamiltonian $H$. It
splits naturally in two parts 
\begin{eqnarray}
  L(t)=L_0+L_1'(t).
  \label{L}
\end{eqnarray}
The first part $L_0=\{\cdots, \, H_0\}$ describes dynamics of the bath in
the potential field of a fixed Brownian particle,
\begin{eqnarray}
  L_0
%=\sum_i\left(
%\frac{\partial H_0}{\partial {\bf p}_i}\cdot
%  \frac{\partial}{\partial {\bf r}_i}
%  -\frac{\partial H_0}{\partial {\bf r}_i} \cdot
%      \frac{\partial}{\partial {\bf p}_i} \right)
  =\sum_i\left(
  \frac{ {\bf p}_i}{m}\cdot \frac{\partial}{\partial {\bf r}_i}
  +{\bf f}_i\cdot \frac{\partial}{\partial {\bf p}_i}
  \right),
\label{L0}  
\end{eqnarray}
where ${\bf f}_i=-\partial (V+\Phi)/\partial {\bf r}_i$ is  a force
on $i$th bath molecule. The second part involves derivatives with respect to
the particle's variables, 
\begin{eqnarray}
  L_1'(t)
%=\frac{\partial H}{\partial P}\frac{\partial}{\partial X}
%  -\frac{\partial H}{\partial X}
%      \frac{\partial}{\partial P}
  =\frac{P}{M}\,\frac{\partial}{\partial X}
  +[F+F_{ex}(t)]\,\frac{\partial}{\partial P},
\end{eqnarray}
where $F=-\partial \Phi/\partial X$ 
and
 $F_{ex}=-\partial U_{ex}/\partial X$
are the $x$-projection of the  forces
exerted on the particle by the bath and external field, respectively.
We temporarily denote $L_1'$ with a prime because another form for this term 
will be introduced shortly.

The standard assumptions and settings of the microscopic theory of Brownian motion
are assumed to be hold. 
Initial conditions for
dynamical variables of the bath are random and distributed according to the 
equilibrium canonical distribution with 
the Hamiltonian $H_0$, 
inverse temperature $\beta=1/(k_B T)$,  
and partition function $Z_0$,
\begin{eqnarray}
  \rho_0=Z_0^{-1}\, \exp(-\beta\, H_0),\quad Z_0=
  \int \exp(-\beta\, H_0)\,\prod_i d{\bf r}_i\,d{\bf p}_i.
\label{rho}
\end{eqnarray}
When averaged over distribution (\ref{rho}), 
the force 
${\bf F}_\nu=-\partial\phi_\nu/\partial {\bf R}_\nu$
from the bath on atom $\nu$ 
vanishes, 
$\langle {\bf F_\nu}\rangle=
\int \rho_0\,{\bf F}_\nu\,\prod_i d{\bf r}_i\,d{\bf p}_i=0$, 
and so does the total force ${\bf F}=\sum_\nu {\bf F}_\nu$ on the cluster, 
$\langle {\bf F}\rangle=\sum_\nu \langle {\bf F}_\nu\rangle =0$.
In order to describe the constrained motion of 
a rigid cluster we need only 
the $x$-projection of the total force ${\bf F}$,
denoted above by $F$ and referred from now on simply as "force", which of
course is also zero on average,
$\langle F\rangle=0$,
whether the cluster is symmetric or not.

Next, the particle's momentum 
is expected to be close to the equilibrium value $\sqrt{M/\beta}$, which is 
$\sqrt{M/m}$ time larger than the equilibrium value of
the bath molecule's momentum $p_e=\sqrt{m/\beta}$.
Then it is convenient 
to work with a 
scaled momentum of the particle
\begin{eqnarray}
P_*=\lambda\, P, \qquad \lambda=\sqrt{m/M}\ll 1
\label{scaling}
\end{eqnarray}
which on average is of the same order of magnitude as $p_e$. When written in
terms of $P_*$, the Liouville operator
takes the form
\begin{eqnarray}
L(t)=L_0+\lambda\,L_1(t),
\label{L2}
\end{eqnarray}
where $L_0$ is still  given by (\ref{L0}) while $L_1=\lambda^{-1}\,L_1'$  reads
\begin{eqnarray}
  L_1(t)=\frac{P_*}{m}\,\frac{\partial}{\partial X}
  +[F+F_{ex}(t)]\,\frac{\partial}{\partial P_*}.
\label{L1}
\end{eqnarray}
The form (\ref{L2})  is more convenient than (\ref{L})
for developing a proper perturbation technique 
since the dependence on the small parameter $\lambda$
in (\ref{L2}) is explicit.

Starting with the equation of motion for the particle's 
scaled momentum  
\begin{eqnarray}
\dot P_*(t)=\lambda\, F(t)+\lambda\, F_{ex}(t),
\label{eqofmotion}
\end{eqnarray}
our goal is to transform this equation into a Langevin form
by partitioning the term $F(t)$, representing the force exerted on the
particle by the bath, into a  dissipative and fluctuating 
parts. While many steps of the derivation are standard, others are less so.
In order to make the paper self-contained, we shall present the derivation in full.

As a preparation, one notes that the equation of motion 
for the force $\dot F(t)=L(t)\,F(t)$ with the initial condition $F(0)=F$
is equivalent to the integral equation $F(t)=F+\int_0^t d\tau\,L(\tau)\,F(\tau)$,
which can be solved by iteration:
\begin{eqnarray}
  F(t)=
  \left\{1+\int_0^t\!\!d\tau L(\tau)
  +\int_0^t \!\!d\tau_1 \int_0^{\tau_1}\!\! d\tau_2\, 
  L(\tau_1)\,L(\tau_2)+\cdots\right\}\,F.
\label{iteration}
\end{eqnarray}
This may be expressed concisely as 
\begin{eqnarray}
F(t)=\overrightarrow\exp \left( \int_0^t \!\! L(\tau)\,d\tau\right)\, F
\label{F1}
\end{eqnarray}
in terms of the time-ordered exponential propagator
\begin{eqnarray}
  \overrightarrow{\exp}\left( \int_0^t \!\! L(\tau)\,d\tau\right)\equiv
  T_+\left\{
  \sum_{n=0}^\infty 
  \frac{1}{n!}\left(\int_0^t \!\!L(\tau)\, d\tau\right)^n
  \right\}
  =1+\int_0^t\!\!d\tau L(\tau)
  +\int_0^t \!\!d\tau_1 \int_0^{\tau_1}\!\! d\tau_2\, 
  L(\tau_1)\,L(\tau_2)+\cdots
\label{propagator}
\end{eqnarray}
where the time-ordering operator $T_+$ rearranges the product of time-dependent
operators in such a way  that time arguments decrease from left to right.
For example,  $T_+\{L(\tau_1)L(\tau_2)\}$ equals $L(\tau_1)L(\tau_2)$
if $\tau_1>\tau_2$ and $L(\tau_2)L(\tau_1)$ otherwise. As a result, operators earlier
in time act before operators at later times.
We shall also need an operator 
\begin{eqnarray}
  \overleftarrow{\exp}\left(-\int_0^t \!\! L(\tau)\,d\tau\right)\equiv
  T_-\left\{
  \sum_{n=0}^\infty 
  \frac{1}{n!}\left(-\int_0^t \!\! L(\tau)\, d\tau\right)^n
  \right\}
  =1-\int_0^t\!\!d\tau L(\tau)
  +\int_0^t \!\!d\tau_1 \int_0^{\tau_1}\!\! d\tau_2\, 
  L(\tau_2)\,L(\tau_1)-\cdots
\label{anti-propagator}
\end{eqnarray}
where the time-ordering operator $T_-$ makes later times operators to appear not
on the left (as $T_+$ does), but on the right. The second equalities
in (\ref{propagator}) and (\ref{anti-propagator}) can be proved by interchanging 
integration variables, see e.g. Ref.~\cite{Tuckerman}.
Note that in our notations 
an arrow over exponentials
indicates a direction of {\it decreasing} time arguments,
which appears not to be a generally accepted convention in the literature.

Operators (\ref{propagator}) and (\ref{anti-propagator}) commute and are inverse to each other,
\begin{eqnarray}
  \overrightarrow\exp\left( \int_0^t \!\! L(\tau)\,d\tau\right)\,
  \overleftarrow\exp\left( -\int_0^t \!\! L(\tau)\, d\tau\right)=
  \overleftarrow\exp\left( -\int_0^t \!\! L(\tau)\,d\tau\right)\,
  \overrightarrow\exp\left( \int_0^t \!\! L(\tau)\,d\tau\right)=1.
  \label{inverse}
\end{eqnarray}
We shall also need the differentiation properties
\begin{eqnarray}
\frac{d}{dt}\,  
\overrightarrow\exp\left( \int_0^t \!\! L(\tau)\,d\tau \right)=
L(t)\,
\overrightarrow\exp\left( \int_0^t \!\! L(\tau)\,d\tau \right),
\qquad
\frac{d}{dt}\,  
\overleftarrow\exp\left(-\int_0^t \!\! L(\tau)\,d\tau \right)=
-\overleftarrow\exp\left(-\int_0^t \!\! L(\tau)\,d\tau\right)\,L(t),
\label{diff}
\end{eqnarray}
which follow directly from the second equalities in 
(\ref{propagator}) and (\ref{anti-propagator}).
 
In the next section we shall use 
expression (\ref{F1}) as a starting point to 
project out the bath variables from the force $F(t)$ 
with a projection operator and perturbation techniques. 
The necessity of
time-ordered exponentials is dictated, of course, by the 
non-commuting of Liouville operators $L(t)$ at different times.
We shall see, however, that up to order $\lambda^3$ (which is sufficient
for our purposes) the time dependent term $F_{ex}(t)\frac{\partial}{\partial P_*}$
in the Liouville operator does not actually affect 
the structure of the dissipative force in the Langevin equation. Knowing that
in advance, one could neglect the time dependence of $L(t)$ and work with,
instead of (\ref{F1}), with the much simpler expression $F(t)=e^{Lt}\,F$.
Up to order $\lambda^3$, such an  ad-hoc simplified approach gives a correct
partitioning of the bath-induced force $F(t)$. It is hard to see, however,
another way to justify this insight but working out (as we do below) 
the exact expression (\ref{F1}).

\section{Nonlinear Langevin equation}

%Note that 
%not all types of ratchets 
%can operate arbitrary close 
%to equilibrium~\cite{Parrondo}.

In this section we exploit
a projection operator technique in the form 
originally developed in Ref.~\cite{KO} for a point-like Brownian particle and to order  $\lambda^2$.
Applying the approach to a rigid asymmetric cluster
of point-like atoms and extending a  perturbation procedure to order $\lambda^3$ we shall be able to derive microscopically the nonlinear Langevin equation (\ref{LE}), which suffices to describe the operation of intrinsic ratchets.

As mentioned above, the idea is to start with the 
exact expression (\ref{F1}) for the force $F(t)$
exerted on the particle by the bath and to partition it into a dissipative (depending only on the particle's momentum)  and fluctuating  parts. To this end,
we need a generalization of the familiar operator identity
\begin{eqnarray}
e^{(A+B)t}=e^{At}+\int_0^t e^{A(t-\tau)}\,B\,e^{(A+B)\tau}\,d\tau
\label{identity11}
\end{eqnarray}
to the case when operators $A$ and $B$ are time-dependent and do not self-commute at different times.
Such generalization has the form
\begin{eqnarray}
&&\qquad\qquad\qquad\qquad \overrightarrow\exp\left(\int_0^t\![A(s)+B(s)]\,ds \right)=
\overrightarrow\exp\left(\int_0^t \!\! A(s)\,ds\right)\nonumber\\
&&+\int_0^t
\overrightarrow\exp\left(\int_0^t \!\! A(s)\,ds \right)\,\,
\overleftarrow\exp\left(-\int_0^{\tau} \!\! A(s)\,ds \right)\,\,
B(\tau)\,\,
\overrightarrow\exp\left(\int_0^\tau \! [A(s)+B(s)]\,ds \right)\, d\tau.
\label{identity1}
\end{eqnarray}
This identity can be verified by multiplying both sides from the left by $\overleftarrow\exp\left(-\int_0^t  A(s)\,ds \right)$
and then differentiating with respect to $t$, also taking into account properties (\ref{inverse}) and (\ref{diff}).

Using (\ref{identity1}) with   
$A=L(t)$ and $B=-\mathcal P\,L(t)$ (with yet an arbitrary operator $\mathcal P$), the force
$F(t)=\overrightarrow\exp\left( \int_0^t  L(s)\,ds \right)\, F$ can be expressed
as
\begin{eqnarray}
F(t)=F_*(t)+\overrightarrow\exp\left(\int_0^t \!\! L(s)\,ds \right)\,
\int_0^t
\overleftarrow\exp\left(- \int_0^\tau \!\! L(s)\,ds \right)\,
\mathcal P \,L(\tau)\,F_*(\tau)\,d\tau
\label{F3}
\end{eqnarray}
where 
\begin{eqnarray}
F_*(t)=\overrightarrow\exp\left( \int_0^t \!\! \mathcal Q\, L(s)\,ds \right)\, F, \qquad \mathcal Q=1-\mathcal P.
\label{F*}
\end{eqnarray}
With a properly chosen operator $\mathcal P$,
expression (\ref{F3}) will eventually represent the desirable partition of $F(t)$, with 
the first term $F_*(t)$ playing the role of the fluctuating Langevin force, while the second term will develop into a dissipative force.

We define $\mathcal P$ as an operator of
averaging (of an arbitrary  dynamical variable $A$) over initial values of bath variables ${\bf r}=\{{\bf r}_i\}$,
${\bf p}=\{{\bf p}_i\}$
with
the canonical distribution (\ref{rho}),
\begin{eqnarray}
\mathcal P\,A=\langle A\rangle=\int \rho_0\,A\,d{\bf r} \,d{\bf p}.
\label{projector}
\end{eqnarray}
Then both $\mathcal P$ and $\mathcal Q=1-\mathcal P$ are projection operators,
$\mathcal P\mathcal P={\mathcal P}$, $\mathcal Q\mathcal Q={\mathcal Q}$, and are orthogonal
\begin{eqnarray}
  \mathcal P\mathcal Q=\mathcal Q\mathcal P=0.
  \label{orthogonal}
\end{eqnarray}
The vanishing of the equilibrium average force exerted by the bath on the particle can now be expressed in the form
\begin{eqnarray}
\langle F\rangle={\mathcal P}\,F=0.
\label{Fiszero2}
\end{eqnarray}
As follows from (\ref{F*})-(\ref{Fiszero2}), the fluctuating component $F_*(t)$
is zero-centered too, 
\begin{eqnarray}
\langle F_*(t)\rangle={\mathcal P}\,F_*(t)=0.
\label{Fiszero3}
\end{eqnarray}
The major benefit of choosing $\mathcal P$ in the form (\ref{projector}) comes from  the relation
\begin{eqnarray}
\mathcal P L_0=\int \rho_0\,L_0\,(...)\,d{\bf r} \,d{\bf p}=0
\label{PL0}
\end{eqnarray}
where the Liouville operator of the bath $L_0$ is given by (\ref{L0}).
%This can be verified by integration by parts  and taking into account that
%$L_0\rho_0\sim \{e^{-\beta H_0}, \,H_0\}=0$. 
This allows one to eliminate the explicit dependence on bath variables in the second term in (\ref{F3}):
\begin{eqnarray}
  {\mathcal P}L(t)F_*(t)={\mathcal P}\,[L_0+\lambda L_1(t)]\,F_*(t)
  =\lambda\,{\mathcal P}\, L_1(t)\, F_*(t)=
  \lambda\,\frac{P_*}{m}\, {\mathcal P}\,
  \frac{\partial}{\partial X}\, F_*(t)+
  \lambda\,\frac{\partial}{\partial P_*}
  {\mathcal P}\, [F+F_{ex}(t)]\,F_*(t).
\label{aux1}
\end{eqnarray}
Moreover, since $\mathcal P\, F_{ex}(t)\, F_*(t)=F_{ex}(t)\, \mathcal P\, F_*(t)=0$,  the dependence on the external force is eliminated as well, 
\begin{eqnarray}
  {\mathcal P}L(t)F_*(t)=
  \lambda\,\frac{P_*}{m}\, {\mathcal P}\,
  \frac{\partial}{\partial X}\, F_*(t)+
  \lambda\,\frac{\partial}{\partial P_*}
  {\mathcal P}\, F\,F_*(t).
\label{aux11}
\end{eqnarray}
Here the first term on the right hand side can be worked out with a useful relation
\begin{eqnarray}
  \frac{\partial}{\partial X}\,\mathcal P\,F_*(t)=0=
  \beta\,\mathcal P\, F\,F_*(t)+\mathcal P\,\frac{\partial}{\partial X}\,F_*(t)
\label{useful_relation}
\end{eqnarray}
to get
\begin{eqnarray}
  {\mathcal P}L(t)F_*(t)=
  \lambda\,\left(
  \frac{\partial}{\partial P_*}-\frac{\beta P_*}{m}\right)\,\mathcal P\,F\,F_*(t)=
  \lambda\,\left(
  \frac{\partial}{\partial P_*}-\frac{\beta P_*}{m}\right)\,\langle F\,F_*(t)\rangle.
\label{aux111}
\end{eqnarray}
Substitution of this into (\ref{F3}) gives for the force  exerted by the bath the following, and still exact,  expression
\begin{eqnarray}
  F(t)=F_*(t)+\lambda\,\overrightarrow\exp\left(\int_0^t \!\! L(s)\,ds \right)\,
\int_0^t
\overleftarrow\exp\left(- \int_0^\tau \!\! L(s)\,ds \right)\,
\left(
  \frac{\partial}{\partial P_*}-\frac{\beta P_*}{m}\right)\,\langle F\,F_*(\tau)\rangle
\,d\tau.
\label{F33}
\end{eqnarray}

%While the external force $F_{ex}$ does not appear here explicitly, it is involved in the expressions
%for the propagators. Therefore the way $F_{ex}$ %would enters the resulting Langevin equation is
%not obvious.

The next step is to expand  
the fluctuating force 
\begin{eqnarray}
F_*(t)=\overrightarrow\exp\left( \int_0^t \!\! \mathcal Q\, L(s)\,ds \right)\, F =
\overrightarrow\exp\left( \int_0^t \!\! [L_0+\lambda\,\mathcal Q L_1(s)]\,ds \right)\, F
\label{F*M}
\end{eqnarray}
in powers of $\lambda$. The first two terms 
will suffice our purpose,
\begin{eqnarray}
  F_*(t)=F_0(t)+\lambda\,F_1(t)+O(\lambda^2).
  \label{expansion}
\end{eqnarray}
Applying iteratively identity (\ref{identity1}) with $A=L_0$ and $B(t)=\lambda\mathcal Q L_1(t)$ to (\ref{F*M}), one gets
\begin{eqnarray}
F_0(t)=e^{L_0 t}\,F, \qquad
F_1(t)=\int_0^t e^{L_0 (t-\tau)}\,
\mathcal Q\, L_1(\tau)\, F_0(\tau)\,d\tau.  
\label{F01}
\end{eqnarray}
The term $F_0(t)$ has a meaning of the force exerted by the bath
on a fixed (or infinitely heavy) particle and does not depend on the particle's momentum $P_*$.  
On the other hand,
the term  $F_1(t)$ originates from the  particle's motion and depends on $P_*$.
This dependence must be explicitly extracted 
in order to work out the second term in (\ref{F33}).

In view of expansion (\ref{expansion}), 
to first order in $\lambda$ 
the force (\ref{F33}) 
acquires the form 
\begin{eqnarray}
  F(t)=F_*(t)-\frac{\lambda\beta}{m}\,\overrightarrow\exp\left(\int_0^t \!\! L(s)\,ds \right)\,
\int_0^t
\overleftarrow\exp\left(- \int_0^\tau \!\! L(s)\,ds \right)\,
P_*\,\langle F\,F_0(\tau)\rangle
\,d\tau.
\label{F333}
\end{eqnarray}
For a low density bath, one can neglect
the coupling of the force on the particle and  the slow hydrodynamic modes of the bath~\cite{Mazo} and to apply the Markovian ansatz 
$\langle F\,F_0(t)\rangle\to\delta(t)\,\int_0^\infty \langle F\,F_0(t)\rangle\,dt$.
Then 
expression (\ref{F333}) 
is further simplified to 
\begin{eqnarray}
  F(t)=F_*(t)-\frac{\lambda\beta}{m}\,P_*(t)
  \int_0^\infty \langle F\,F_0(t)\rangle\,dt.
\label{F3333}
\end{eqnarray}
Substitution of this into the equation of motion
(\ref{eqofmotion}) yields the linear Langevin equation 
\begin{eqnarray}
   \dot P_*(t)=-\lambda^2\,\zeta_0 \,P_*(t)+\lambda\,F_*(t)+\lambda\,F_{ex}(t),
\label{LE_lin2}
\end{eqnarray}
with
the dissipating coefficient
\begin{eqnarray}
  \zeta_0=\frac{\beta}{m}\,\int_0^\infty \langle F\,F_0(t)\rangle \,dt.
  \label{z0}
\end{eqnarray}
In terms of the unscaled momentum $P=P_*/\lambda$, Eq. (\ref{LE_lin2}) takes
the standard form (\ref{LE_lin})
\begin{eqnarray}
   \dot P(t)=-\gamma_0 \,P(t)+F_*(t)+F_{ex}(t),
\label{LE_lin2M}
\end{eqnarray}
with the dissipating coefficient
\begin{eqnarray}
  \gamma_0=\lambda^2\,\zeta_0=\frac{\beta}{M}\,\int_0^\infty \langle F\,F_0(t)\rangle \,dt
  \label{gamma0}
\end{eqnarray}
and zero-centered fluctuating force, $\langle F_*(t)\rangle=0$.
Thus in lowest order in $\lambda$  the Langevin equation for an asymmetric rigid cluster
has the same form as for a point-like Brownian particle and cannot account for the
operation of intrinsic ratchets.

More interesting, and in fact sufficient for our purpose, is an approximation of the exact expression (\ref{F33}) for $F(t)$ to order $\lambda^2$. Since the second term in (\ref{F33}) contains the  factor $\lambda$, it suffices to substitute there a linear approximation
$F_*(t)=F_0(t)+\lambda\, F_1(t)$.
The component $F_1(t)$ is given by (\ref{F01}), or more explicitly
\begin{eqnarray}
  F_1(t)&=&\int_0^t d\tau\, e^{L_0(t-\tau)}\mathcal Q\, \left(
  \frac{P_*}{m}\,\frac{\partial}{\partial X}+[F+F_{ex}(t)]\,
  \frac{\partial}{\partial P_*}
  \right)
    \,F_0(\tau).
\end{eqnarray}
Since $F_0(t)$ does not depend on $P_*$, this is reduced to
\begin{eqnarray}
  F_1(t)=\frac{P_*}{m}\int_0^t d\tau\,\left(
  e^{L_0(t-\tau)}\,\frac{\partial}{\partial X}\,F_0(\tau)
  -\mathcal P \frac{\partial}{\partial X}\,F_0(\tau)
  \right).
  \label{F_1B}
\end{eqnarray}
Here, similar to (\ref{useful_relation}), 
$
\mathcal P \frac{\partial}{\partial X}\,F_0(t)=-\beta
\langle F\,F_0(t)\rangle
$, which gives
\begin{eqnarray}
  F_1(t)=\frac{P_*}{m}\int_0^t d\tau\,\left(
  e^{L_0(t-\tau)}\,\frac{\partial}{\partial X}\, F_0(\tau)
  +\beta\,\langle F\,F_0(\tau)\rangle
  \right).
  \label{F_1C}
\end{eqnarray}
Note that 
to first order in $\lambda$ the fluctuating force
$F_*(t)\approx F_0(t)+\lambda\,F_1(t)$ is independent of  the external force
$F_{ex}(t)$ (this is not so for higher perturbation orders).

% Does $F_2(t)$  depend on $F_{ex}$?
% I think it does (for an asymmetric particle)!

With expression (\ref{F_1C}) 
for $F_1(t)$ at hand, 
we can  
work out
the correlation function 
$
  \langle F\,F_*(t)\rangle=
  \langle F\,F_0(t)\rangle+\lambda\,\langle F\,F_1(t)\rangle
$
in (\ref{F33}) as follows:
\begin{eqnarray}
  \langle F\,F_*(t)\rangle=C_0(t)+\lambda\,\frac{P_*}{m}\,C_1(t),
  \label{corr3}
\end{eqnarray}
where
\begin{eqnarray}
  C_0(t)=\langle F\,F_0(t)\rangle,\qquad
  C_1(t)=
  \int_0^t d\tau\,\left\langle
  F\,
  e^{L_0(t-\tau)}\,\frac{\partial}{\partial X}\,F_0(\tau)\right\rangle.
  \label{C01}
\end{eqnarray}
It can be proved with a symmetry argument (see the Appendix) that
for a symmetric particle in  a uniform bath the correlation $C_1(t)$
vanishes identically 
(in that case the first non-zero correction to the
weak-coupling approximation $\langle F\,F_*(t)\rangle=C_0(t)$
is of order $\lambda^2$). 
On the other hand, for an asymmetric particle
the correlation $C_1(t)$ does not vanish in general and ultimately is responsible for  the operation of intrinsic ratchets.

Substitution of (\ref{corr3}) into (\ref{F33}) yields  for $F(t)$ an approximation of order $\lambda^2$
\begin{eqnarray}
  F(t)=F_*(t)-\frac{\lambda\beta}{m}\,\overrightarrow\exp\left(\int_0^t \!\! L(s)\,ds \right)\,
\int_0^t
\overleftarrow\exp\left(- \int_0^\tau \!\! L(s)\,ds \right)\,
\left\{
C_0(\tau)\,P_*+\frac{\lambda}{m}\,C_1(\tau)\,\left(
P_*^2-\frac{m}{\beta}\right)
\right\}
d\tau.
\label{F4}
\end{eqnarray}
In the Markovian approximation 
$C_i(t)\to \delta(t)\,\int_0^\infty C_i(t)\,dt$, $i=0,1$,
this expression  acquires the form
\begin{eqnarray}
  F(t)=F_*(t)-\frac{\lambda\beta}{m}\,P_*(t)
 \int_0^\infty\!\!\! C_0(\tau)\,d\tau
 -\frac{\lambda^2\beta}{m^2}\,
\left(
P_*^2(t)-\frac{m}{\beta}\right)
\int_0^\infty \!\!\! C_1(\tau)\,d\tau.
\label{F44}
\end{eqnarray}
Finally, substitution of this expression into the equation of motion (\ref{eqofmotion}) yields
the nonlinear Langevin equation of order $\lambda^3$
\begin{eqnarray}
  \dot P_*(t)=-\lambda^2\,\zeta_0\,P_*(t)
  -\lambda^3\,\zeta_1\,\big( P_*^2(t)-\langle P_*^2\rangle_e\big)
  +\lambda\,F_*(t)+\lambda\,F_{ex}(t),
  \label{LE2}
\end{eqnarray}
where $\langle P_*^2\rangle_e=m/\beta$ is the equilibrium value of the
scaled momentum squared, the dissipative constants are
\begin{eqnarray}
  \zeta_0=\frac{\beta}{m}\,\int_0^\infty C_0(t)\,dt, \qquad
  \zeta_1=\frac{\beta}{m^2}\,\int_0^\infty C_1(t)\,dt,
\label{FDT}
\end{eqnarray}
the correlation functions $C_0(t)$ and $C_1(t)$ are given
by (\ref{C01}), and $F_*(t)$ is a zero-centered noise, 
$\langle F_*(t)\rangle=0$.
Adopting the Markovian approximation we assume
that the characteristic time $\tau_0$  of correlations
$C_0(t)$ and $C_1(t)$ is distinctly shorter than both
the period of the external force $F_{ex}(t)$ and the momentum relaxation time $\tau_p=1/\lambda^2\zeta_0$. Then Eq. (\ref{LE2}) corresponds to a coarse-grain description with a time resolution 
$\Delta t\gg \tau_0$.

%There is some subtlety in justifying the Markovian ansatz
%beyond the lowest order in $\lambda$~\cite{P}, yet to order %$\lambda^3$
%the results (\ref{LE3}) and (\ref{FDT}) can still be %recovered.

In terms of the unscaled momentum $P=P_*/\lambda$,
the Langevin equation (\ref{LE2}) has the form (\ref{LE}), 
\begin{eqnarray}
  \dot P(t)=-\gamma_0\,P(t)
  -\gamma_1\,\big(P^2(t)-\langle P^2\rangle_e\big)
  +F_*(t)+F_{ex}(t),
  \label{LE3}
\end{eqnarray}
where $\langle P^2\rangle_e=M/\beta$ is the equilibrium value of the
square momentum, and dependence on $\lambda$ is now absorbed
in the rescaled dissipation coefficients
\begin{eqnarray}
  \gamma_0=\lambda^2\zeta_0=
  \frac{\beta}{M}\,\int_0^\infty C_0(t) \,dt,\qquad
  \gamma_1=\lambda^4\zeta_1=
   \frac{\beta}{M^2}\,\int_0^\infty C_1(t) \,dt.
\label{gammas}
\end{eqnarray}
Relations (\ref{FDT}) or (\ref{gammas}) are ought to be viewed as
fluctuation-dissipation relations for an asymmetric Brownian particle. 
Note again that 
for a symmetric particle the correlation function $C_1(t)$ and the nonlinear 
dissipation coefficient $\gamma_1$ vanish and the first
nonlinear correction is of order $\lambda^4$ and cubic in the momentum~\cite{Mori,PF}.

In the next section we shall prefer to evaluate the particle's drift  using the 
Langevin equation for the scaled momentum 
in the form (\ref{LE2}), which involves the small parameter $\lambda$ explicitly.

\section{Evaluation of drift}
Let us derive a set of two coupled equations for the first two moments of 
the particle's scaled momentum
\begin{eqnarray}
  A(t)=\langle P_*(t)\rangle,
  \qquad
  B(t)=\langle P_*^2(t)\rangle-\langle P_*^2 \rangle_e.
\end{eqnarray}
As in the previous section, the angular 
brackets denote averaging over initial values of bath variables with 
the distribution (\ref{rho}). 
Taking average of the Langevin equation (\ref{LE2}) and assuming that 
the external force $F_{ex}(t)$ is
uniform (does not depend on the particle's position), 
one obtains  the first equation
\begin{eqnarray}
  \dot A(t)=-\lambda^2\,\zeta_0\,A(t)
  -\lambda^3\,\zeta_1\,B(t)+\lambda\, F_{ex}(t).
  \label{A}
\end{eqnarray}
A systematic directional motion of the particle is expected to 
emerge as a result of the coupling of $A(t)$
and $B(t)$. Since the latter
appears in (\ref{A}) multiplied by $\lambda^3$,
it is sufficient to complement (\ref{A}) with a familiar equation for the second 
moment to order  $\lambda^2$,
\begin{eqnarray}
  \dot B(t)=-2\,\lambda^2\zeta_0\,B(t)
  +2\lambda\,F_{ex}(t)\,A(t).
  \label{B}
\end{eqnarray}
The easiest way to obtain this equation is multiplying the 
linear Langevin equation
(\ref{LE_lin2}) by $2P_*(t)$, taking average, and  
taking into account that $\langle P_*(t)\,F_0(t)\rangle=
\lambda\,\zeta_0\,m/\beta= \lambda\,\zeta_0\,\langle P_*^2\rangle_e$
The latter relation can be  directly verified using an explicit
solution $P_*(t)$
of the linear equation (\ref{LE_lin2}), the Markovian anzatz for the autocorrelation of $F_0(t)$,
and the fluctuation-dissipation relation (\ref{z0}).

Eqs. (\ref{A}) and (\ref{B}) form a closed system
of linear equations for the first two moments of $P_*$.
They are equivalent to Eqs. (\ref{moments1}) and (\ref{moments2}) of the Introduction and  
to those exploited in Ref.~\cite{Broeck}. 
Since we are looking for a stationary solution, the specific choice of initial conditions is immaterial.
Assuming for simplicity
$A(0)=B(0)=0$, the solution of (\ref{A}) can be written as
\begin{eqnarray}
  A(t)=\lambda\,a_1(t)+\lambda^3\,a_2(t)
\label{A1}
\end{eqnarray}
where
\begin{eqnarray}
  a_1(t)=\int_0^t dt'\,e^{-\lambda^2\zeta_0(t-t')}\,F_{ex}(t'),
  \quad
  a_2(t)=-\zeta_1 \int_0^t dt'\,e^{-\lambda^2\zeta_0(t-t')}\,B(t'),
\label{A2}
\end{eqnarray}
and the  solution of (\ref{B}) is 
\begin{eqnarray}
  B(t)=2\,\lambda\int_0^t dt'\,e^{-2\lambda^2\zeta_0(t-t')}\,F_{ex}(t')\,A(t').
\label{B2}
\end{eqnarray}
Equations (\ref{A1})  and (\ref{B2}) can be uncoupled by substituting in (\ref{B2}) an approximation 
$A(t)=\lambda\,a_1(t)+\lambda^3\,a_2(t)\approx\lambda\,a_1(t)$ which gives 
\begin{eqnarray}
  B(t)=2\lambda^2\int_0^t dt' e^{-2\lambda^2\zeta_0(t-t')}\,F_{ex}(t')
  \int_0^{t'} dt'' e^{-\lambda^2\zeta_0(t'-t'')}\,F_{ex}(t'').
\label{A2_1}
\end{eqnarray}
With this expression for $B(t)$, the second equation of
(\ref{A2}) yields for $a_2(t)$
\begin{eqnarray}
a_2(t)=-2\lambda^2\,\zeta_1\,\int_0^t dt_1\,e^{-\lambda^2\zeta_0(t-t_1)}
  \int_0^{t_1} dt_2\,e^{-2\lambda^2\zeta_0(t_1-t_2)}\,F_{ex}(t_2)
\int_0^{t_2} dt_3\,e^{-\lambda\zeta_0(t_2-t_3)}\,F_{ex}(t_3).
\label{a2}
\end{eqnarray}
As a result, for the average scaled momentum
$\langle P_*\rangle=A=\lambda\,a_1+\lambda^3\,a_2$ one gets
\begin{eqnarray}
  \langle P_*(t)\rangle&=&
  \lambda\,\int_0^t dt_1\,e^{-\lambda^2\zeta_0(t-t_1)}\,F_{ex}(t_1)\nonumber\\
&-& 2\,\lambda^5\,\zeta_1\,\int_0^t dt_1\,e^{-\lambda^2\zeta_0(t-t_1)}
  \int_0^{t_1} dt_2\,e^{-2\lambda^2\zeta_0(t_1-t_2)}\,F_{ex}(t_2)
\int_0^{t_2} dt_3\,e^{-\lambda^2\zeta_0(t_2-t_3)}\,F_{ex}(t_3).
\label{P_scaled}
\end{eqnarray}

This result is obtained under the assumption that the second term in 
(\ref{A1}) or (\ref{P_scaled}) is much smaller than the first one. Contrary to what
expressions (\ref{A1}) or (\ref{P_scaled}) may suggest, the smallness of $\lambda$ 
alone does not guarantee the validity of this approximation 
since both terms may actually be of the same order in $\lambda$; this  will be shown 
below explicitly for specific forms of $F_{ex}(t)$. On the other hand, 
it is clear from (\ref{P_scaled}) that the approximation does  hold for 
sufficiently small values of $\zeta_1$ and/or amplitude of the external force. 
The precise consistency condition will be formulated shortly.

Let us re-write the above result for the  unscaled momentum $P=\lambda^{-1}P_*$ 
as a sum of two contributions,
\begin{gather}
\langle P(t)\rangle= \langle P(t)\rangle_0+\langle P(t)\rangle_1,
\label{P}\\
\langle P(t)\rangle_0=\int_0^t dt_1\,e^{-\gamma_0(t-t_1)}\,F_{ex}(t_1),
\label{P_lin}\\
\langle P(t)\rangle_1=- 2\,\gamma_1\,\int_0^t dt_1\,e^{-\gamma_0(t-t_1)}
  \int_0^{t_1} dt_2\,e^{-2\gamma_0(t_1-t_2)}\,F_{ex}(t_2)
\int_0^{t_2} dt_3\,e^{-\gamma_0(t_2-t_3)}\,F_{ex}(t_3),
\label{P_nonlin}
\end{gather}
where we  use
the rescaled dissipation coefficients 
$\gamma_0=\lambda^2\,\zeta_0$ and
$\gamma_1=\lambda^4\,\zeta_1$ as they 
appear in the Langevin equation (\ref{LE3}) for $P$ and given explicitly by (\ref{gammas}).
Here the first term $\langle P(t)\rangle_0$, representing  the linear response to the external force,
can  be obtained from
the linear Langevin equation (\ref{LE_lin2M}).
It contributes to the particle's net drift only
if $F_{ex}(t)$ is biased,
i.e. when the  time-average of $F_{ex}(t)$ is non-zero.
The second term 
$\langle P(t)\rangle_1$
represents  a nonlinear response contribution originating from
the coupling of $\langle P\rangle$ and $\langle P^2\rangle$ to higher order in
$\lambda$. Being quadratic in the external force $F_{ex}(t)$, this term
can produce a directional drift of the particle even when $F_{ex}(t)$ is unbiased. 
The drift may be characterized by the net momentum $P_{net}$, or net
velocity $V_{net}=P_{net}/M$,
defined as a time-average, which we shall denoted by an overbar, of the ensemble average 
$\langle P(t)\rangle$,
\begin{eqnarray}
  P_{net}=\overline{\langle P(t)\rangle}=
  \lim_{T\to\infty} \frac{1}{T}\int_0^T  \langle P(t)\rangle \, dt.
\label{P_net}
\end{eqnarray}

Consider first a harmonic external force $F_{ex}(t)=F_0\,\sin\omega t$. 
In this case, from  (\ref{P})-(\ref{P_net})
one finds for $t\gg1/\gamma_0$
\begin{eqnarray}
  P_{net}=-\frac{\gamma_1}{2\,\gamma_0^3}\,\,
  \frac{F_0^2}{1+(\omega/\gamma_0)^2}.
\label{P_net_harmonic}
\end{eqnarray}
This expression originates
entirely from the nonlinear response term $\langle P(t)\rangle_1$ given by 
(\ref{P_nonlin}), $P_{net}=\overline{\langle P(t)\rangle}_1$.
The linear response term 
has a form
\begin{eqnarray}
\langle P(t)\rangle_0=\frac{F_0}{\gamma_0^2+\omega^2}\,
  \left\{
  \gamma_0\,\sin\omega t+\omega\,(e^{-\gamma_0t}-\cos\omega t)
  \right\}
  \label{P_lin_harm}
\end{eqnarray}  
and vanishes after averaging over time in the long-time limit, $\overline{\langle P(t)\rangle}_0=0$.

Recall that we obtained the results (\ref{P}) and (\ref{P_net_harmonic}) 
under the assumption that the nonlinear contribution to the average momentum 
is much smaller than the linear one,
$\langle P(t)\rangle_1\ll \langle P(t)\rangle_0$. According to (\ref{gammas}),
the dissipation coefficients scale with the particle's mass as
\begin{eqnarray}
  \gamma_0\sim M^{-1}, \qquad \gamma_1\sim M^{-2}.
  \label{M}
 \end{eqnarray}
Then it follows from (\ref{P_net_harmonic})-(\ref{M}) that
for the low-frequency domain 
$\omega\ll \gamma_0$, both linear 
$\langle P(t)\rangle_0$ and 
nonlinear $\langle P(t)\rangle_1\sim P_{net}$ contributions scale in the same 
way (linearly) with $M$. As was already noted, this means 
that the condition of small mass ratio, $\lambda\ll 1$, is not sufficient for 
the consistency of the above approach.
A comparison of Eqs. (\ref{P_net_harmonic}) and (\ref{P_lin_harm}) shows that 
one has to require an additional condition
\begin{eqnarray}
  \frac{\gamma_1}{\gamma_0^2}\,F_0
  \ll 1,\qquad \mbox{for} \quad \omega\ll \gamma_0.
  \label{constraint1}
\end{eqnarray}
According to (\ref{M}), this constraint does not involve the particle's mass $M$, 
and implies  a small value of the nonlinear dissipation coefficient $\gamma_1$ 
and/or of the external force amplitude $F_0$. On the other hand, for 
the high-frequency domain instead of (\ref{constraint1}) one gets
\begin{eqnarray}
  \frac{\gamma_1}{\gamma_0\, \omega}\,F_0
  \ll 1,\qquad \mbox{for} \quad \omega\gg \gamma_0.
  \label{constraint2}
\end{eqnarray}
Since $\gamma_1/\gamma_0\sim M^{-1}$, this  condition can be satisfied 
for sufficiently small $\lambda$.

The quantity $\gamma_0$ has a meaning of
the inverse time for the
particle's momentum relaxation, $\gamma_0=1/\tau_p$,
and for a micro-meter sized
Brownian object in water has a value of order about $10^7\,s^{-1}$.
%($10^4\,s^{-1}$ in the air).
Then according to (\ref{P_net_harmonic}) the frequency-independent maximum value of the net momentum 
\begin{eqnarray}
P_{max}=-\frac{\gamma_1}{2\,\gamma_0^3}\, F_0^2
\label{P_max}
\end{eqnarray}
corresponds to the frequency of long-radio waves. Remarkably, according to (\ref{M}) and
(\ref{P_max}), the maximum  net velocity $V_{max}=P_{max}/M$ does not depend on the particle's mass.

%The results of this section can be readily extended 
%for $F_{ex}(t)$ being an arbitrary periodic function of time, expanding
%$F_{ex}(t)$ in Fourier series.

Consider also the case when  the external
force is  a stationary random processes
$F_{ex}(t)=F_{ex}(t\,|\,\theta)$ whose specific realizations
depends on a random parameter (or a set of parameters)  $\theta$.
One may think of  $F_{ex}(t\,|\,\theta)$, for instance,
as of a force generated by randomly moving external charges 
with a set $\theta$ of initial coordinates and momenta.  
Whatever physical system is responsible for generating the force
$F_{ex}(t\,|\,\theta)$,
we shall assume that it is not in equilibrium
with the thermal bath, and that the average of $F_{ex}(t\,|\,\theta)$  over $\theta$, 
which we denote as $\langle F_{ex}(t)\rangle_\theta$, is zero.

For this case,  the momentum averaged over the bath $\langle P(t)\rangle$
is still given by expressions (\ref{P})-(\ref{P_nonlin}),
but now we define the net momentum $P_{net}$ as an average of $\langle P(t)\rangle$ over
$\theta$, rather than over time. Again, since $\langle F_{ex}(t)\rangle_\theta=0$ 
the linear response $\langle P(t)\rangle_0$ does not contribute to the net momentum, while 
taking the average over $\theta$ of the expression (\ref{P_nonlin}) 
for the nonlinear response $\langle P(t)\rangle_1$ gives
\begin{eqnarray}
  P_{net}=
  -2\,\gamma_1\,\int_0^t dt_1\,e^{-\gamma_0(t-t_1)}
  \int_0^{t_1} dt_2\,e^{-2\gamma_0(t_1-t_2)}
  \int_0^{t_2} dt_3\,e^{-\gamma_0(t_2-t_3)}\,
  \langle F_{ex}(t_2)\,F_{ex}(t_3)\rangle_\theta.
\label{P_net_noise}
\end{eqnarray}
Suppose the  auto-correlation function of the fluctuating external force $F_{ex}(t)$ is exponential,
\begin{eqnarray}
  \langle F_{ex}(t)F_{ex}(t')\rangle_\theta=F_0^2\,\exp\left(
  -\frac{|t-t'|}{\tau_c}
  \right).
\label{noise_corr}
\end{eqnarray}
Then for $t\gg 1/\gamma_0$ the net momentum (\ref{P_net_noise})
acquires  the form
\begin{eqnarray}
  P_{net}=-\frac{\gamma_1}{\gamma_0^3}\,
  \frac{F_0^2}{1+1/(\gamma_0\,\tau_c)}.
\label{P_net_noise2}
\end{eqnarray}
In this expression, the noise  correlation time $\tau_c$ plays a role similar to the inverse frequency 
of a harmonic force, and the validity conditions are similar to (\ref{constraint1}) and (\ref{constraint2}). 
As a function of $\tau_c$, the net momentum reaches the maximal value
in the limit of $\gamma_0\, \tau_c\gg 1$, i.e. when the correlation time of
$F_{ex}(t)$ is much longer than the momentum
relaxation time $\tau_p=1/\gamma_0$.

%Let us compare the net momenta of a dimer activated by a noisy and harmonic
%external forces assuming that the correlation time for the former and the
%period of the latter are of the same order of magnitude
%$\omega\,\tau_c\sim 1$. In this case,
%for a high frequency $\omega/\gamma_0\gg 1$ and
%short correlation time $\gamma_0\,\tau_c\ll 1$ one finds from
%(\ref{P_net_noise2}) and (\ref{P_net_harmonic}) that the ratio of $P_{net}$
%for the two types of the external force is of order
%\begin{eqnarray}
%  \frac{(P_{net})_{noise}}{(P_{net})_{harmonic}}\sim \frac{\omega}{\gamma_0}\gg 1.
%\end{eqnarray}
%Thus, the net directional motion activated by a short-correlated external
%noise is much larger than that activated by a high-frequency harmonic force.
%Perhaps looking strange at the thirst glance, this observation is actually
%intuitively plausible considering that a noisy  force has  
%fragments which takes much longer than $\tau_c$ to change a sign,
%while for a deterministic harmonic force such time is fixed.

\section{Simulation}
In this section  we describe molecular dynamics simulations of intrinsic ratchets based on
two- and three-atom clusters
immersed in the two-dimensional ideal gas.
Our goal here is to provide simple illustrations, rather than quantitative verification
of theoretical predictions of the previous sections.
Such a verification would require an explicit theoretical evaluation of
dissipation coefficients $\gamma_0$ and $\gamma_1$ with relations (\ref{gammas}),
which is beyond the scope of the present study.
Instead, the simulation may be used to get some empirical insight about how $\gamma_0$ and $\gamma_1$ depend on a cluster's structure. 
%and also to test qualitative predictions of the theory.
%We shall focus on  the latter and  make no attempts
%to mimic any realistic physical systems. 

While the theoretical model discussed in the previous sections assumes
that the ratchet's atoms are connected by rigid bonds,
in our simulations we consider clusters of atoms connected by stiff harmonic bonds.
We found that transport properties of ratchets only weakly depend on the value of 
the bond strength constant $k$, at least when the latter is  sufficiently large. 
Therefore simulation results for stiff clusters ($k$ is large but finite) 
and theoretical predictions for their rigid prototypes ($k\to \infty$) are expected to be close.

%The direct simulation of such systems
%is often infeasible, requiring a very small time increment $\Delta t$
%(smaller than the inverse bond frequency). 
%On the other hand, in our case
%we found that the cluster's drift is practically independent on $k$,
%and therefore is expected to be the same as for a rigid system, 
%already for  moderate values of $k$ for which direct simulation
%is still feasable.

\subsection{Dimer activated by harmonic force}  Consider
a diatomic cluster (dimer) consisting of two atoms of the same mass $M_a=M/2$, 
oriented and constrained to move
along the $x$-axis. Each atom is subjected
to the external harmonic drive
$F_{ex}(t)=F_0\,\sin\omega t$
applied in the $x$-direction.
The two atoms interact with each other via  a harmonic potential
\begin{eqnarray}
U=\frac{k}{2}\,(X_2-X_1-d)^2,
\label{potential_harmonic}
\end{eqnarray}
where $d$ is the length of the unstretched dimer,
$X_1$, $X_2$ are positions of left and right atoms, respectively.
The parameters $k$ and $d$ are assumed to be sufficiently large
to guarantee  that  $X_2>X_1$  at any time.

Atom $\nu$ ($\nu=1,2$) interacts with  molecule $i$
of the surrounding two-dimensional bath through truncated
repulsive potential of the  form
\begin{eqnarray}
  \phi_\nu({\bf R}_\nu, {\bf r}_i)=\frac{u_\nu}{\alpha_\nu}\,\left(
  \frac{|{\bf R}_\nu-{\bf r}_i|}{\sigma_\nu}
  \right)^{-\alpha_\nu}\,h(\sigma_\nu-|{\bf R}_\nu-{\bf r}_i|),
\label{repulsive}
\end{eqnarray}
where ${\bf R}_\nu=(X_\nu,0)$ and  ${\bf r}_i=(x_i,y_i)$
are position vectors of 
atom $\nu$ and  molecule $i$, respectively, $\sigma_\nu$
are the interaction radius for atom $\nu$, and $h(x)$
is the Heaviside step function. The latter makes
the force corresponding to potential (\ref{repulsive})
to be  zero when the distance $|{\bf R}_\nu-{\bf r}_i|$
between a molecule and atom exceeds $\sigma_\nu$.
An artificial singularity of the force at  $|{\bf R}_\nu-{\bf r}_i|=\sigma_\nu$
is of no consequence for our purposes. 

\begin{figure*}[t]
\includegraphics[height=4cm]{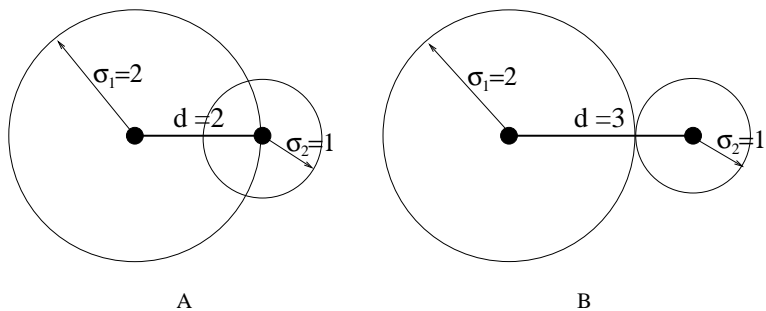}
\caption{Two dimer configurations with the same
  radii of the atom-molecule interaction spheres, $\sigma_1=2$  and $\sigma_2=1$,
  and different lengths $d$. Configuration $A$ with $d=2$
  (on the left) is characterized by a significant overlapping of the
  interaction spheres and is found to have a
  larger value of the net drift velocity.}
\end{figure*}

\begin{figure*}[t]
\includegraphics[height=7cm]{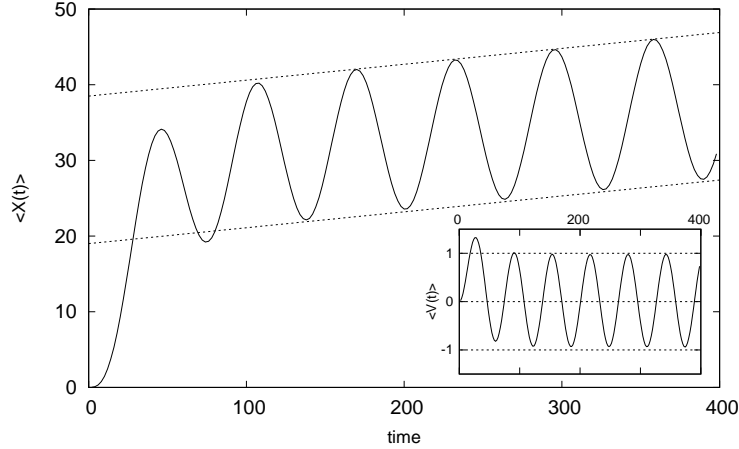}
\caption{Main plot: The average displacement $\langle X(t)\rangle$
  of a dimer of configuration $A$
  (left on Fig. 2) activated by the  external harmonic force
  $F_{ex}(t)=F_0\sin\omega t$ (acting on each atom) with 
  $F_0=2$ and $\omega=0.1$. For units and values of other parameters see
  Eq. (\ref{units}) and the text above it.
  The slope of tangent lines (dashed lines) to minima and maxima
  gives the net velocity of the dimer $V_{net}\approx 0.02$.
  Inset: The corresponding average velocity $\langle V(t)\rangle$.
 The amplitude of positive peaks of the curve
  $\langle V(t)\rangle$ is slightly larger than that of negative ones,
  which results in  the net motion to the right. The net velocity
  $V_{net}$ can be also determined as a time average of
  $\langle V(t)\rangle$.
  }
\end{figure*}

In this setting, the dimer's asymmetry may be due to 
unequal potential parameters for the two atoms.
As a specific  example,  we consider dimer configurations with the potential
radius of the left atom to be two times larger
than  that of the right one,
while other parameters of the potentials for both atoms are the same, 
\begin{eqnarray}
  \sigma_1=2\sigma_2, \quad u_1=u_2=u, \quad \alpha_1=\alpha_2=\alpha.
\label{config}
\end{eqnarray}
Two such configurations, denoted as $A$ and $B$,
which differ  by the dimer's length $d$ are shown in Fig. 2.
The simulation shows that configuration $A$ 
with $d=\sigma_1=2\sigma_2$ 
develops a larger net velocity
than configuration $B$
with $d=1.5\sigma_1=3\sigma_2$. 
In fact, it turns out
that for a low-density thermal bath configuration $A$ shows a maximum net velocity among all other
configurations of type (\ref{config}) with different lengths $d$. 
%(Check: this statement might be correct only for a given bath density. Is config A  optimal
%for other values of the bath density?)
The main plot in Fig. 3 shows simulation results for the average
center-of-mass displacement  $\langle X(t)\rangle$
of a dimer of configuration $A$ subjected to a harmonic drive
$F_{ex}(t)=F_0\sin\omega t$. The slope of a tangent lines to maxima or minima of 
the curve  $\langle X(t)\rangle$ equals to the net velocity of the dimer.
Alternatively, the drift may be visualized 
with a plot of the average center-of-mass velocity $\langle V(t)\rangle$ shown in the inset of Fig. 3.
One may notice that maxima of the
curve $\langle V(t)\rangle$ have slightly larger amplitudes than
minima, which results in a positive net velocity.
According to theoretical result (\ref{P_net_harmonic}), which we rewrite here for the net velocity in the form
\begin{eqnarray}
  V_{net}=
  \frac{V_{max}}{1+(\omega/\gamma_0)^2},\qquad
  V_{max}=-\frac{\gamma_1}{2\,M\,\gamma_0^3}\,F_0^2,
\label{V_net_harmonic}
\end{eqnarray}
a positive net velocity 
corresponds to a negative value of the nonlinear dissipation coefficient, $\gamma_1<0$.

The simulation was performed for the following set of parameters:
 molecule-atom mass ratio $m/M_a=0.05$ (which corresponds to  $\lambda^2=m/M=m/(2M_a)=0.025$),
% molecule-atom mass ratio $m/M_a=0.05
concentration of bath molecules $\rho=0.2$,
interaction exponent $\alpha=6$,
energy interaction coefficient $u=1$,
squared bond frequency $\Omega^2=k/M_a=1$,
external force frequency $\omega=0.1$ and amplitude $F_0=2$.
Here and below we adopt
the following units of length $x_0$,
velocity $v_0$, time $t_0$, energy $u_0$, and force $f_0$:
\begin{eqnarray}
  x_0=\sigma_2, \quad v_0=v_e, \quad
  t_0=\sigma_2/v_e, \quad u_0=m\,v_e^2,\quad f_0=m\,v_0/t_0
\label{units}
\end{eqnarray} 
where $v_e=1/\sqrt{m \beta }$ denotes the average thermal speed  of a molecule 
of the bath in equilibrium.

The data presented in Fig. 3 give for
configuration $A$ with $d=2$ the net velocity value approximately
$V_{net}=0.02$. This is about one order of magnitude smaller than the equilibrium thermal speed of the dimer 
$V_e=1/\sqrt{M\beta}$, which in given units equals to
$\lambda=\sqrt{m/M}\approx 0.16$.
For configuration $B$ with $d=3$ the simulation under the same conditions 
shows the drift  about two times slower, with
$V_{net}\approx 0.01$.
For configurations with $d>4$
and the same set of other parameters the net drift is getting
very small and difficult to detect. On the other hand, decreasing $d$ from the apparently optimal 
value $d=2$ also results in fast decreasing of the net velocity.
While theoretical result (\ref{V_net_harmonic})
for  $V_{net}$ involves two parameters $\gamma_0$ and $\gamma_1$,
we found that $\gamma_0$ depends on $d$ only weakly.
Then  the above-mentioned dependence of $V_{net}$ on $d$ should be attribute 
mostly to that of $\gamma_1$ and the underlying
correlation $C_1(t)$.
%see (\ref{gammas}) and (\ref{C01}). 

With $V_{net}$ found from the simulation and the linear dissipation coefficient $\gamma_0$
evaluated independently (by simulating relaxation of the average  momentum 
$\langle P(t)\rangle=P(0)\,e^{-\gamma_0\,t}$ in the absence of an external force), one can use 
Eq. (\ref{V_net_harmonic}) to evaluate the maximum  net velocity $V_{max}$ and the nonlinear 
dissipation coefficient $\gamma_1$. Let us estimate the order of the magnitude of the  former:
We found $\gamma_0\approx 0.03$ 
for configuration $A$ and a slightly larger (by about $10\%$) value
for configuration $B$. For the external force
frequency $\omega=0.1$ adopted in our simulation, we have 
$(\omega/\gamma_0)^2\sim 10$ for both configurations.
Then, according  to
(\ref{V_net_harmonic}),
$V_{net}\sim 0.1\, V_{max}$. 
On the other hand, the simulation gives
$V_{net}\sim 0.1\, V_e$, and we conclude that $V_{max}$ is  comparable
to the equilibrium thermal speed $V_e$ of the dimer. A similar estimation $V_{max}\sim V_e$ was found in Ref.~\cite{Broeck} for  mesoscopic ratchets of cone shapes
and in Ref.~\cite{granular1,granular2} for granular ratchets.

A word of warning is in order here.
The value $F_0=2$ for the external force's amplitude, adopted in our simulation,
was chosen sufficiently large to make the drift easily noticeable on the plot of
$\langle X(t)\rangle$ in Fig. 3. This value is likely to be too high and beyond the validity
range of the theory. The latter 
assumes that the particle's velocity is close to its
equilibrium value, $V\sim\lambda \, v_e$. In our simulation units (\ref{units}),
this assumption reads $V\sim \lambda\ll 1$ and, as the inset in Fig. 3 shows, for $F_0=2$
it is clearly violated. Therefore, if one wishes to use similar
simulations for an accurate estimation of  $\gamma_1$ and $F_{max}$,  a much weaker external field
should probably be employed.

%Simulations with different $\omega$ show 
%the frequency dependence of the net motion consistent
%with (\ref{P_net_harmonic}).

\subsection{Dimer activated by external telegraph noise}
Let the external force $F_{ex}(t)$ be  a stochastic dichotomous
Markov process (also known as a telegraph noise) flipping between two states $\pm F_0$ with  
a constant transition  rate $k$, see the inset in Fig. 4.
The life-time $\tau$ of each of two states, i.e. the waiting time
between two successive flips, is a random
variable with the  exponential distribution $f(\tau)=k\,\exp(-k\,\tau)$.
In this case, $F_{ex}(t)$ has 
an exponential autocorrelation function of  the form (\ref{noise_corr}) with
the correlation time $\tau_c=1/(2k)$, and a theoretical result for the net momentum
is given by (\ref{P_net_noise2}).

\begin{figure*}[t]
\includegraphics[height=6cm]{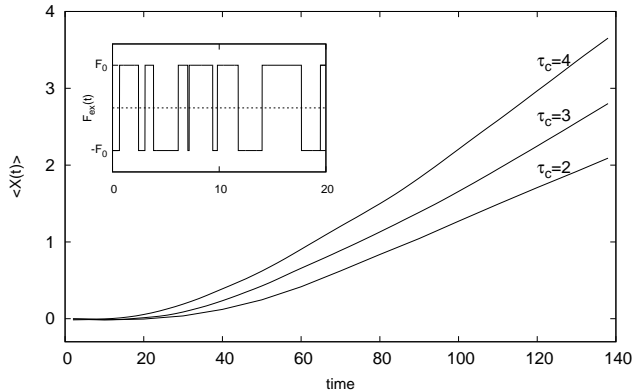}
\caption{Main plot: The average displacement $\langle X(t)\rangle$ of a  dimer of   
configuration $A$ (left on Fig. 2) activated by the external telegraph noise $F_{ex}(t)$ 
for three values of the correlation time $\tau_c$. Inset: A specific realization of $F_{ex}(t)$ with $\tau_c=1$. 
The average $\langle X(t)\rangle$ is calculated over about $10^5$ simulation runs.
}
\end{figure*}

Fig. 4 shows simulation results for the  displacement of the dimer of
configuration A (see Fig. 2), double-averaged over initial values of bath variables and over
realizations of $F_{ex}(t)$, for several values of the correlation
time $\tau_c$.  Other parameters
are the same as for the simulation described in the previous subsection. 
With $\gamma_0\approx 0.03$ and $\tau_c\sim 1$, the parameter $\gamma_0\tau_c$ is small,
$\gamma_0\,\tau_c\ll 1$. For this regime, 
Eq. (\ref{P_net_noise2}) predicts that the net velocity can be approximated as 
$V_{net}\approx -\gamma_1\, \tau_c\, F_0^2/(M\,\gamma_0^2)$, increasing linearly with $\tau_c$.
The data presented in Fig. 4 are qualitatively consistent with this dependence, and can be used 
to estimate $\gamma_1$.   Compared to a harmonic external  drive, such estimation may be more laborious 
since a  much larger number of simulation runs is required
in order to average out fluctuations 
(still visible on Fig. 4) 
of the curves $\langle X(t)\rangle$ and $\langle V(t)\rangle$.

%This is, of course, because for the a fluctuating %$F_{ex}(t)$
%the average is taken not only over the bath but also over realizations of
%$F_{ex}(t)$. 

\subsection{Trimer activated by harmonic force}
Consider a cluster of three atoms (trimer) in the shape (when undisturbed)
of isosceles triangle with the base $2a$, altitude $h$, and oriented
with the altitude parallel to the $x$-axis, see Fig. 5.
Atoms of the cluster are constrained to move (without friction)
along the $x$-axis only, each along its own rail.
Atoms are connected by harmonic springs with the same force 
constant $k$, and the internal potential energy of the cluster is
\begin{eqnarray}
U=\frac{k}{2}\,(l_{12}-l_{12}^{0})^2+
\frac{k}{2}\,(l_{13}-l_{13}^{0})^2+
\frac{k}{2}\,(l_{23}-l_{23}^{0})^2,
\end{eqnarray}
where  $l_{ij}$ are the
distances between atoms $i$ and $j$ as functions of their instantaneous
$x$-coordinates,
\begin{eqnarray}
  l_{12}=\sqrt{(X_1-X_2)^2+a^2}, \quad l_{13}=\sqrt{(X_1-X_3)^2+a^2},\quad
  l_{23}=\sqrt{(X_2-X_3)^2+4a^2},
\end{eqnarray}
and $l_{ij}^0$ and the corresponding distances
in mechanical equilibrium
\begin{eqnarray}
  l_{12}^0=l_{13}^0=\sqrt{a^2+h^2}, \quad l_{23}^0=2a.
\end{eqnarray}
A value of the string constant $k$ is chosen large enough to preserve
the trimer's orientation,
i.e. to guarantee that at all time the vertex atom $1$ is
at the right of the base atoms $2$ and $3$, $X_1>X_2, X_3$.

\begin{figure*}[t]
\includegraphics[height=6cm]{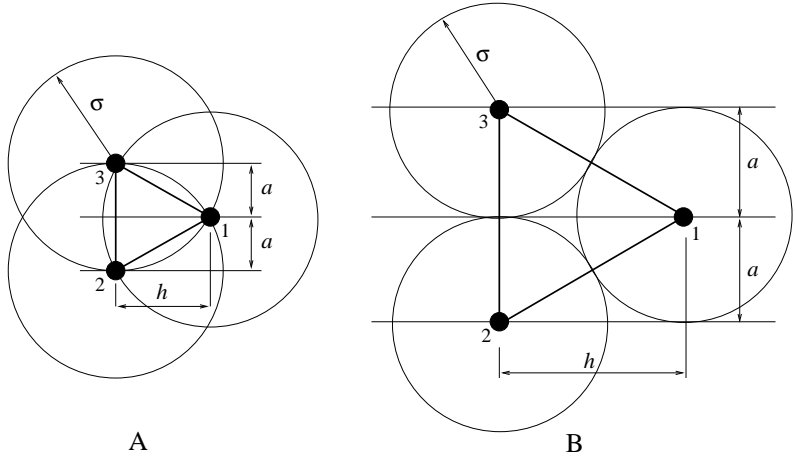}
\caption{ Two equilateral trimer configurations (in equilibrium) with the same radius $\sigma$  
of bath-atom interaction spheres  and different side lengths $2a$. The simulation  shows
that the net drift velocity of configuration $A$ with $2a=\sigma$ (on the left) 
is more than three times  higher than that of 
configuration $B$ with $a=\sigma$ (on the right), see Fig. 6.}
\end{figure*}

Each of three atoms $\nu=1,2,3$ interacts with a bath molecule $i$
by repulsive  potential (\ref{repulsive}) of the same strength $u$,
exponent $\alpha$, and radius $\sigma$,
\begin{eqnarray}
  \phi_\nu({\bf R}_\nu, {\bf r}_i)=\frac{u}{\alpha}\,\left(
  \frac{|{\bf R}_\nu-{\bf r}_i|}{\sigma}
  \right)^{-\alpha}\,h(\sigma-|{\bf R}_\nu-{\bf r}_i|),
\label{repulsive2}
\end{eqnarray}
where the position vectors of atoms are
${\bf R}_1=(X_1,0)$, ${\bf R}_2=(X_2,a)$,  ${\bf R}_3=(X_3,-a)$,
and ${\bf r}_i=(x_i,y_i)$ are position vectors of bath molecules.
Atoms have the same mass
$M_a$, and the total mass $M=3M_a$ of the cluster
is much larger than that of a bath molecule $m$. Each atom
is subjected by the same external harmonic force
$F_{ex}(t)=F_0\,\sin\omega t$
applied in the $x$ direction.
The simulation units are still given by (\ref{units}), with $\sigma$ as a unit of length.

\begin{figure*}[t]
\includegraphics[height=6cm]{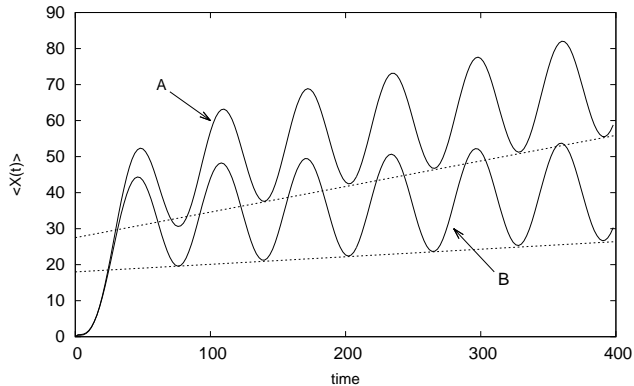}
\caption{ The average displacement (solid lines) 
of two trimer configurations $A$ and $B$ shown 
in Fig. 5 activated by a harmonic external force. 
The slopes of tangent lines to minima (dashed) give the net  velocity $V_{net}$ of a  cluster. We found 
$V_{net}\approx0.07$ for configuration $A$ and $V_{net}\approx 0.02$ 
for configuration $B$.
The simulation parameters are: 
the molecule-atom mass ratio $m/M_a=0.075$ (molecule-trimer mass ratio $m/M=0.025$), 
external force frequency $\omega=0.1$ and amplitude $F_0=2$,
squared harmonic bond frequency $\Omega^2=k/M_a=4.5$,  
bath density $\rho=0.2$. 
%Units are defined by Eq. (\ref{units}).
} 
\end{figure*}

The simulation shows that 
the trimer's net velocity  depends
on the atomic geometry of the cluster, that is on the ratio
of the altitude $h$ and the base $2a$.
The maximum drift was found 
for the equilateral trimer ($h/a=\sqrt{3}$). 
On the other hand, for a given trimer's atomic geometry the drift
strongly depends on the ratio of a characteristic geometric length, say $a$,
and  the radius $\sigma$ of the atom-molecule interaction sphere.
For an equilateral trimer, the simulation suggests that  the maximal mobility
is achieved for configuration $A$ with  $\sigma=2a$, see Fig. 5.

What is characteristic for this maximum drift configuration 
is an optimal combination
of a strong overlapping of atom-molecule interaction spheres, 
and still significant asymmetry of the spheres' union.
Increasing of the $a$ with $\sigma$ kept fixed would enhance the asymmetry of the spheres' 
union but decrease their overlapping, as for configuration $B$ in Fig. 5. 
Such a configuration is characterized by a smaller drift velocity, see Fig. 6.
On the other hand, increasing the interaction spheres radius $\sigma$ with atomic 
cluster size $a$ kept constant would increase the spheres' overlapping but reduce
the asymmetry of the spheres' union. Again, this results in a decrease of the cluster's drift velocity.  
These trends corroborate those we observed in the previous subsections for dimers and, 
according to (\ref{P_net_harmonic}), reflect dependence of the nonlinear dissipation 
coefficient $\gamma_1$ and the underlying correlation $C_1(t)$ on structural properties of the cluster.

%We use units similar to as before,  see (\ref{units}), except that 
%now the unit length is  $\sigma$,

%Project:
%The total force is larger than for dimer because
%it act on the additional atom. May be good idea
%to get data for trimer with the same total force
%and plot data together  for dimer and trimer
%when ALL parameters are same (including $\Omega$) 

\section{Conclusion}
In this paper we considered intrinsic ratchets as microscopic clusters of atoms interacting with 
bath molecules through given short-range potentials. 
The assumption that clusters atoms are connected by rigid bonds, adopted in 
the theoretical part of this paper, does not appear to be  restrictive. The nonlinear Langevin equation  for 
soft clusters can be derived in a similar way 
eliminating  internal degrees of freedom of a cluster with a properly modified  projector operator.

Compared to previous studies, 
our approach describes ratchets with broader types of asymmetry and also emphasizes 
the  generality of the relevant nonlinear Langevin equation (\ref{LE3})  
and fluctuation-dissipation relations (\ref{C01}) and (\ref{gammas}), particularly for 
the nonlinear dissipation coefficient $\gamma_1$. The latter 
is a key quantity which determines
the drift velocity of a ratchet.
%and is expressed as an integral of the correlation function
%$c_1(t,\tau)=\langle F e^{L_0 %(t-\tau)}(\partial/\partial X) %F_0(\tau)\rangle$.
While fluctuation-dissipation relations
play important role in the theory, they are difficult to use and rarely exploited
for a direct evaluation of dissipation coefficients. Instead, 
the linear dissipation coefficient $\gamma_0$ can be readily estimated experimentally or in 
simulation as the inverse momentum relaxation time, and the 
the nonlinear dissipation coefficient 
$\gamma_1$ can be determined comparing
a measured or simulated net velocity $V_{net}$ with theoretical relations for $V_{net}$ obtained in this paper.
%The theory itself can be verified by showing that such evaluated $\gamma_1$ has close values for different types of the external force $F_{ex}(t)$. 
%This would require, however, more accurate and systematic simulations than that presented in this paper.

Our results suggest that a value of $\gamma_1$  and the underlying correlation function 
$C_1(t)$ depend in a delicate way on the composition of atom-molecule interaction
spheres,  rather than of the mere geometry of the cluster's atomic skeleton. 
For a symmetric cluster or a single atom $\gamma_1$ is zero, but
it may be negligibly small for asymmetric clusters as well.
On the one hand, $\gamma_1$ tends to increase for a larger overlapping of interaction spheres.
On the other hand, the increase of the spheres' overlapping with their radius kept 
fixed diminishes the asymmetry of the spheres' union. The maximum value of $\gamma_1$ 
(and therefore a cluster's maximum drift velocity) is achieved  
for an optimal combination of the two factors, as for configurations $A$ on Figs. (2) and (5).

Let us also note a subtle role of the external force $F_{ex}(t)$ in the operation of intrinsic ratchets.
On the one hand, this force "shakes" the system preventing it from reaching
thermal equilibrium. For $F_{ex}(t)=F_0\sin\omega t$, the efficiency of this "shaking" is expected 
to increase with $\omega$. On the other hand, the particle's drift during one half-period of the force
oscillation decreases with $\omega$. A nontrivial interplay of these two  factors is 
that the drift 
velocity $V_{net}$ increases when $\omega$ decreases and, according 
to the presented theory,  
reaches a maximum in the adiabatic regime, i.e. when $F_{ex}(t)$
varies infinitely slow. Such behaviour is in contrast with 
many other types of ratchets
activated by a modulated external potential, which  
cannot work arbitrary close to equilibrium~\cite{review1,review2,Parrondo}.
Our simulations confirm qualitatively  the theoretical dependence of $V_{net}$ on $\omega$ 
(and on the correlation time $\tau_c$, for a fluctuating external force), though the adiabatic 
regime is of course  directly inaccessible in simulation.

The nonlinear Langevin equation obtained in this paper is of 
the same form as for the familiar problem of an adiabatic piston separating two gases 
of different temperatures~\cite{PS2}. 
The adiabatic piston problem can be formulated in a one-dimensional form and under certain model 
assumptions allows an analytical evaluation of relevant correlation functions and dissipation coefficients~\cite{PS2}. 
Then it might be tempting to develop a
more simple model of intrinsic ratchets
using a one-dimensional geometry of the piston problem. 
For instance, one may assume that the 
left and right sides of the piston are made of different materials, so that molecules of left and right 
gases (now at the same temperature) interact with the piston via different potentials. One might expect that a piston with such structural asymmetry would behave as an intrinsic ratchet, i.e.
develop an average drift velocity
when subjected to an external unbiased 
low-frequency force. 
Remarkably, we found no evidence for that. 
For a specific model 
where the thermal bath is a uniform  ideal gas whose molecules
interact with the piston via a parabolic potential (of different amplitudes for left and right piston's sides), 
analytical results of Ref.~\cite{PS2} predict that $\gamma_1=0$ and therefore there is no drift. 
Our simulations show no drift for other types of asymmetric potentials as well. 
This is perhaps not surprising
considering the two sides of the piston as analogues of the dimer's atoms and 
recalling the importance 
of the overlapping of atom-molecule interaction regions as a condition for the drift. 
There is no such overlapping for the piston geometry, and therefore there is no drift.

Instead of the nonlinear Langevin equation, one can use an equivalent
corresponding master equation for the velocity or momentum distribution function
$f(P,t)$. To order $\lambda^3$ the master equation
differ from  the standard second order Fokker-Planck equation by involving the $P$-derivative of order three~\cite{VKO,Kampen,Meurs,Plyukhin_physa}.
A common worry is  that according 
the Pawula theorem~\cite{Risken}  a master equation of order higher than two may not  preserve
positivity of the distribution function. 
To answer this, it was emphasized  elsewhere~\cite{positivity} that  
an {\it approximate} distribution function 
does not need to be 
positive for all values of its arguments. 
It may take a small negative values in far-tail regions 
and still be computationally useful.
The terms with higher order derivatives are 
also of higher order in $\lambda$, and as long as one treats them consistently as perturbations, the results are 
meaningful and valid to a given order in $\lambda$. 
Several model calculations showed that 
an approximate distribution given by master equations of higher orders may be negative in the far-peripheral regions, which are computationally negligible since the distribution there is very small;
on the other hand, for the region(s) where
the distribution is large
the corrections stemming from the higher derivative terms may be important and do not violate the distribution's positivity ~\cite{Risken,Vollmer,Hasegawa,Grima,Grima2,Cianci}.
In practice, instead of solving a master equation of higher order
it is usually easier to get and solve perturbatively equations for the moments.
In this form, the validity and usefulness of the method 
was demonstrated in many works, e.g. ~\cite{Broeck,PS2,PF,Plyukhin_therm,Meurs,Kawai2,Kawai3,Cianci2}.

% Another type of the dimer's asymmetry
% is unequal masses of the atoms, $M_1\ne M_1$.

  \renewcommand{\theequation}{A\arabic{equation}}
  % redefine the command that creates the equation no.
  \setcounter{equation}{0}  % reset counter 
  \section*{APPENDIX}  % use *-form to suppress numbering

%Q: What happen if the total external force on the particle's %atoms is zero?
%This is the case for a dimer composed of two oppositely %charged ions.
%In the rigid cluster approximation - nothing. But how about
%a flexible cluster?

Let us show that for a symmetric Brownian particle 
consisting of a single atom interacting with molecules of the bath via a spherically symmetric potential, 
the correlation function $C_1(t)$ given by (\ref{C01}), and therefore the nonlinear damping 
coefficient $\gamma_1=(\beta/M^2) \int_0^\infty C_1(t)\, dt$, are zero.
In that case the bath Hamiltonian $H_0$ (\ref{H3}) 
depends on  position  vectors of the particle ${\bf R}=(X,Y,Z)$ and molecules
${\bf r}_i=(x_i,y_i,z_i)$
only through the lengths of the
difference vectors ${\bf q}_i={\bf r}_i-{\bf R}$ and
${\bf q}_i-{\bf q}_{i'}={\bf r}_i-{\bf r}_{i'}$.
Under the inversion of the phase space 
\begin{eqnarray}
 {\bf p}_i\to -{\bf p}_i, \qquad {\bf q}_i\to -{\bf q}_i
\label{inversion}
\end{eqnarray}
the Hamiltonian $H_0$ and the Liouville operator $L_0=\{\cdots, H_0\}$ are
invariant, while the operator 
  $\frac{\partial}{\partial X}=
 \sum_i \frac{X-x_i}{|{\bf q}_i|}\,\frac{\partial}{\partial|{\bf q}_i|}
  $
  is odd,
\begin{eqnarray}
  H_0\to H_0, \quad L_0\to L_0,\quad \frac{\partial}{\partial X}\to
  -\frac{\partial}{\partial X}.
\label{inversion2}
\end{eqnarray}
According to (\ref{C01}),  
$C_1(t,\tau)=\int_0^t \langle A(t,\tau)\rangle\,d\tau$ where
\begin {eqnarray}
  \langle A(t,\tau)\rangle=\frac{1}{Z_0}\int e^{-\beta H_0} A(t,\tau) \,
  \prod_i d{\bf q_i}\,d{\bf p_i}
\label{aux5}
\end{eqnarray}
is the average of the dynamical function
\begin {eqnarray}
A(t,\tau)= F\,
e^{L_0(t-\tau)}\,\frac{\partial}{\partial X}\,F_0(\tau)=
\frac{\partial H_0}{\partial X}\,e^{L_0(t-\tau)}\,
     \frac{\partial}{\partial X}\,e^{L_0\tau}\,
     \frac{\partial H_0}{\partial X}.
\end{eqnarray}
Taking into account (\ref{inversion2}) one notices that under the inversion
transformation (\ref{inversion}) the function $A$ is odd,  $A\to -A$.
Making in
the integral (\ref{aux5}) the substitution ${\bf q}_i=-{\bf q}_i'$ and
${\bf p}_i=-{\bf p}_i'$, and taking into account
the symmetry properties of $H_0$
and $A$, one finds $\langle A(t,\tau)\rangle=-\langle A(t,\tau)\rangle=0$, and 
$C_1(t)=\int_0^t \langle A(t,\tau)\rangle \, d\tau=0$.

%The result that $\gamma_1=0$ is of course well known and means that
%for a point-like Brownian particle the Langevin equation to higher
%orders in $\lambda$ does not contain even powers
%of particle's momentum. 

%In a similar way one can prove the vanishing of
%$c_1(t,\tau)$  for symmetric, i.e. satisfying conditions (\ref{sym_geom}) and
%(\ref{sym_comp}), multi-atom clusters.

%\acknowledgements{
%I thank Richard Bowles, Stephen Shea, and the referees 
%for stimulating comments.}   

%\end{multicols}

\end{document}